\documentclass{article}

\usepackage{arxiv}

\usepackage[utf8]{inputenc} 
\usepackage[T1]{fontenc}    
\usepackage{hyperref}       
\usepackage{url}            
\usepackage{booktabs}       
\usepackage{amsfonts}       
\usepackage{nicefrac}       
\usepackage{microtype}      
\usepackage{cleveref}       
\usepackage{lipsum}         
\usepackage{graphicx}
\usepackage{natbib}
\usepackage{doi}

\usepackage{listings}     
\usepackage{lstautogobble}  
\usepackage{color}          
\usepackage{zi4}            

\definecolor{bluekeywords}{rgb}{0.13, 0.13, 1}
\definecolor{greencomments}{rgb}{0, 0.5, 0}
\definecolor{redstrings}{rgb}{0.9, 0, 0}
\definecolor{graynumbers}{rgb}{0.5, 0.5, 0.5}

\usepackage{listings}
\title{Geo-located data for better dynamic replication}

\date{April, 2022}

\author{Luís M. Silva, Frederico Aleixo, Albert van der Linde, João Leitão, Nuno Preguiça\\
	NOVA LINCS \& DI, FCT, Universidade NOVA de Lisboa\\
	\texttt{\{lmt.silva, fp.aleixo, a.linde\}@campus.fct.unl.pt, \{jc.leitao, nuno.preguica\}@fct.unl.pt} \\
}

\renewcommand{\headeright}{Technical Report}
\renewcommand{\undertitle}{Technical Report}
\renewcommand{\shorttitle}{Geo-located data for better dynamic replication}

\hypersetup{
pdftitle={Geo-located data for better dynamic replication},
pdfsubject={An Revised Version},
pdfauthor={Luís M. Silva, Frederico Aleixo, Albert van der Linde, João Leitão, Nuno Preguiça},
pdfkeywords={Location-based data, Overlay Networks, Partial replication},
}

\begin{document}
\maketitle

\begin{abstract}
An increasing number of mobile applications share location-dependent information, from collaborative applications and social networks to location-based games. 
For such applications, peer-to-peer architectures where mobile devices share information directly may grant lower latency and reduce server load.
In this work, we propose a framework to support these applications, providing location-dependent replication. Data has an associated location and is replicated in the mobile devices that show interest and are close to that location. 
Our proposal combines peer-to-peer synchronisation among mobile devices and replication at edge servers. 
Our preliminary results are promising.
\end{abstract}



\section{Introduction}

As modern applications shift more and more towards a paradigm of significant interaction among users, and as mobile devices become the preferred platform for deploying new applications, supporting data sharing among mobile devices becomes an increasingly important problem. Good examples are multiplayer games, mapping services with real-time crowdsourced traffic and alerts or any hyperlocal data distribution system with content like ads, local news alerts or social media features.
 
Most current applications rely on centralised infrastructures for data storage and coordination. With a fast-expanding number of clients, workloads on the server-side skyrocket and swiftly grow as a scalability bottleneck. Another concern is providing the best user experience. Applications with strict latency requirements, such as multiplayer games or AR/VR content, need the quickest dissemination of information. 

An alternative approach is to rely on peer-to-peer architectures, where clients communicate and share information directly (\cite{legion}). This leads to reduced server load and lower latency. However, these systems typically do not consider that the users' interest set depends on their location and users' movement may change that data set. Furthermore, they usually do not leverage the existence of edge nodes that can help and improve the system's regular operation.

This paper proposes a framework for replicating geo-located data, assuming data objects have an associated location and that clients are interested in data tagged with a position close to theirs, e.g. in Pokémon GO (\cite{pokemongo}), a player is interested in close-by objects and players. 
Our replication protocol explores nodes' hierarchy, covering client nodes running typically on mobile devices, edge, and cloud servers. Client nodes connect directly with neighbouring clients to synchronise their state, keeping replicas of objects with nearby locations. As a client moves, data become irrelevant. Resulting in being discarded and connections to distant clients ended.
Edge nodes replicate objects of nearby locations. The benefit of a hierarchy of edge and server nodes is to help the replication process and enforce data durability, even in the presence of fast node movement and sparse node distribution.

The main contributions of this paper are: 1) the GeoLoc Architecture including an Overlay and Bully algorithms - that manage the creation of hierarchical topologies, including edge resources; and 2) our reasoning on applications that may benefit from our work, complemented by a short evaluation examining and validating our concept.


\section{Related work and Background}

\textbf{Proximity detection failure.}
Neighbour discovery is a necessary and primordial feature in location-based mobile applications.
Applications such as Waze's location-based alerts, ads, and multiplayer games utilize this process to perceive which clients are in the surroundings.
However, this does not translate into efficient interactions between clients, as servers are needed to route messages.
Some works use an agnostic view to the communications channels to offer proximity-based discovery and connections in challenging topologies (\cite{google_nearby}).
Yet, small hyper-local ad hoc networks, like the resulting ones from those mechanisms, do not allow contact with clients in dispersed locations.

\textbf{Communication between clients.}
Communication mechanisms have been studied to iteratively adapt overlay network topology tracking network resources and leverage client connections. These are often concerned about neighbour proximity, link swapping techniques, and building decentralised multicast mechanisms supporting peer-to-peer broadcast (\cite{legion,ganesh2001scamp, voulgaris2005cyclon, leitao2007hyparview, leitao2012x}) .
Our work aims to deal with high churn rates and address locality-aware data support so clients can fetch data from nearby resources relieving the computation cost from distant ones.

\textbf{Emergence of 5G technologies.}
Technologies centred on the 5G framework are still uncertain and not well defined. Small mixed deployments with 4G providing control signalling and 5G for fast data connections emerge globally. Current specifications allow us to expect some features, such as low latency, device-to-device communications, and enabling mobile edge computing (\cite{dolui2017comparison}).
Edge computing-based applications have a place in diverse sectors of the 5G architecture - base stations, cell towers, and small local data centres. Even as an emerging technology, it enables computation and storage closer to clients, releasing network and cloud infrastructure load.

\textbf{Data Sharing at the Edge.}
Communication patterns and network stability tend to be widely diverse on the edge compared to highly mobile peer-to-peer networks due to the absence of standardized edge storage protocols and a general lack of steer by the research community caused by the field's novelty.
Works tend to consider the edge as a CDN for static content or a caching strategy, allowing for better network performance. The type of data and form of caching have been studied, in the form of Popularity-based approaches (\cite{zhang2015edgebuffer}), Mobility-Aware Caching (\cite{wang2017mobility}) and Peer Metadata Sharing (\cite{song2017content}).
We leverage some of these strategies, where we see room for improvement.

\textbf{Cloud Load Balancing.}
Pokémon GO's launch success hugely surpassed expectations; the actual requests per second rate was near 50x their worst-case estimate. The game is highly interactive and communal among users to further complicate the matter. All players in a given area share the same view of the game world and interact with each other inside that location. This requires the production and distribution of near-real-time updates of state shared by all participants. Famously, Niantic's backend platform, hosted at Google Cloud, received so many requests that they could not keep pace with all inbound connections and scale to meet those demands. With the load balancing system exceeding the available capacity, the traffic wave caused a cascading failure scenario. Degradation was observed in numerous services (APIs, datastores, and applicational backends), ultimately causing requests to time out and leaving clients unanswered (\cite{beyer2018site}).


\section{Framework and System Design}

\begin{figure}[t]
\begin{center}
\includegraphics[width=.45\textwidth]{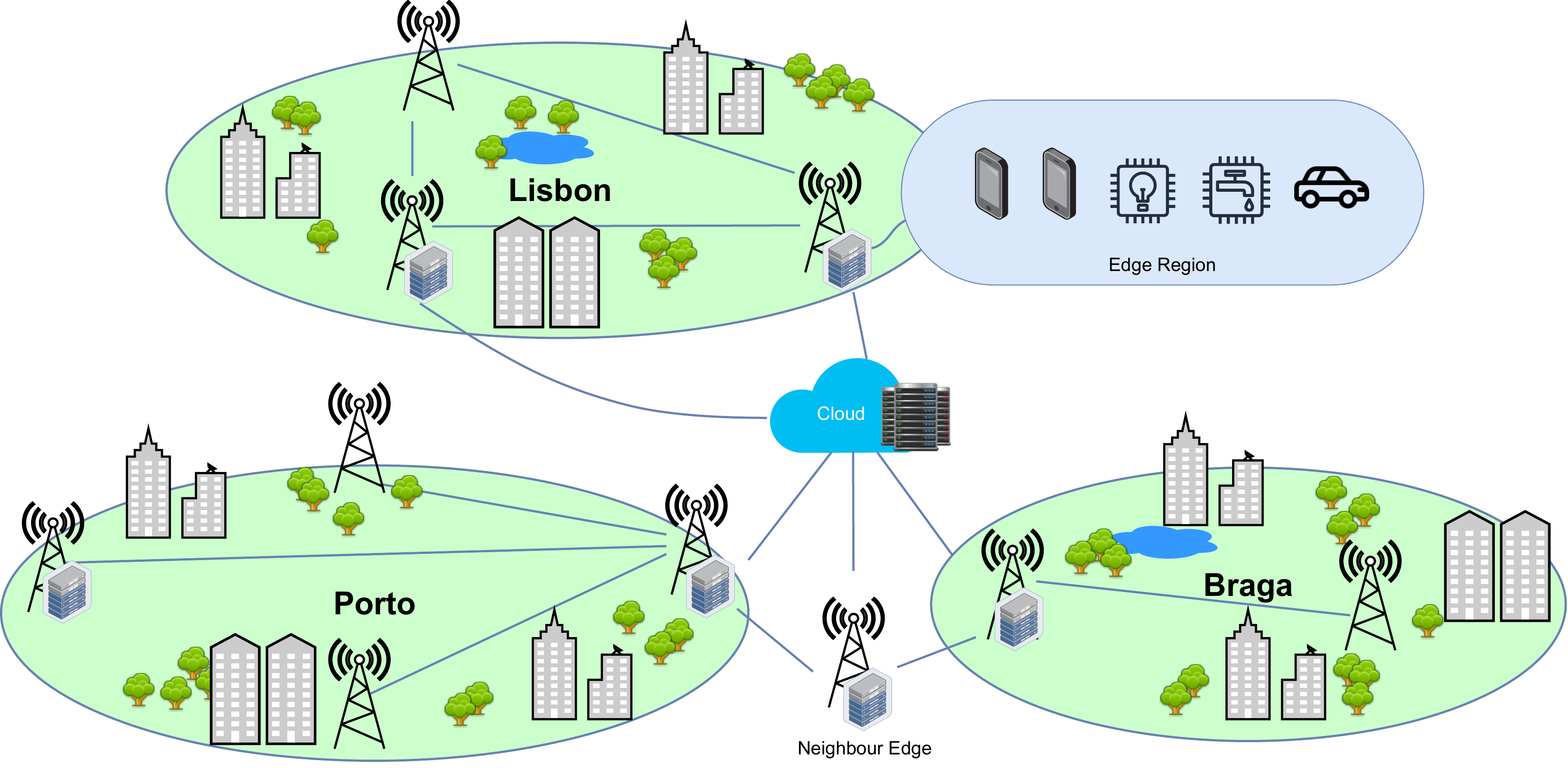}
\caption{Hierarchical System}
\label{fig:hierarchy}
\end{center}
\end{figure}

The GeoLoc framework design assumes a hierarchical system composed of mobile devices, edge servers, and cloud servers, as observed in Fig.~\ref{fig:hierarchy}.
We consider mobile devices as client nodes, standing at the base of the hierarchical system. Communication preferably follows a peer to peer model in an asynchronous fashion between them.
At higher levels of the hierarchy, edge and cloud servers are present -- clients communicate with these servers either directly or through peers (to minimize connections to the servers).
Clients communicate among them over an established infrastructure (cellular, WiFi) or in an ad-hoc fashion (as WiFi ad-hoc or Bluetooth), i.e., our framework is communication-medium agnostic. Still, we consider that most messages with a server as the destination will resort to some infrastructure.
Our architecture includes server components for signalling and object storage. Servers on edge networks can be provisioned as demand dictates and serve a determined geographical region. Higher up, global servers hold all data and facilitate the data dissemination and synchronization processed between regions.
The benefit from this hierarchical construct is to grant applications with the means to swiftly disseminate information among nodes positioned in an edge resource's vicinity.

According to the application logic, nodes and servers alike create and manipulate data objects. Additionally, every data object holds an associated geographical position. Spatial locality is exploited to classify objects as relevant to client nodes -- client nodes hold only the subset of the objects relevant to them, considering their location. Since client nodes move between different locations, the interest in neighbouring objects will follow that path.
Edge nodes replicate data objects tagged with nearby locations. As edge nodes are assumed to be at a fixed spot, the interest set of edge nodes remains mostly stable, simplifying the replication process.
A global identifier distinguishes each node. The system maintains a record of the nodes' current and past geographical coordinates.
Although our system focuses on storing data objects tagged with a location, it also supports objects that might need to be replicated independently of their location.

\textbf{\textit{Architecture.}}
The GeoLoc framework relies on several components facilitating a peer sampling mechanism based on node and data's physical location. First, an Overlay Algorithm is responsible for creating logical communication channels between nearby nodes that share common interests. 
Second, an Objects Bully algorithm ensures that every update made to an object arrives at an object's server - typically an edge server. Updates are propagated through the overlay, reaching the server directly or via peers in the overlay. Nodes in the overlay update the copies of the objects they are interested in.

\textbf{\textit{The GeoLoc Overlay Algorithm.}}
Since the system's focus is on supporting replicating data objects based on their location by a nearby client, we devised an overlay network to help in this purpose. The idea is to keep an overlay where a node connects directly to the closest nodes, where these connections are changed as nodes move.

A signalling server keeps a record of all node positions, providing a mechanism for clients nodes to receive a list of nearby neighbours that can become peers, as seen in Listing~\ref{lst:sigserver}. This policy reduces the need to propagate messages throughout the overlay and solves the proximity detection failure problem.

By default, a means to contact a server is parameterized, with nodes trying to keep an open connection to it at all times. Depending on the location, the clients will be redirected to contact the nearby edge server.
When a new node joins the overlay, its location is sent to a server. In turn, when a server receives a node's position, it always returns with a list of nodes physically close and willing to become peers.

Periodically and when a node moves a certain amount of distance units, it will broadcast to its peers and server its current position, as depicted on Listing~\ref{lst:overlayclient}. That way, nearby peers do not need to wait for the next server update to register this new information. When a peer drifts too far away from a calculated radius, it discards that connection.
Since this protocol thrives for independence from a server, it will still send its new positions to its peers should a node become disconnected from the signalling server. In turn, they acknowledge its location and propagate it to their neighbours. A server disconnection results in clients with a more limited global vision on close-by nodes, but that can still make new connections.

\begin{figure}
\begin{lstlisting}[language=bash, caption=Signalling server logic, label=lst:sigserver]
const MAX_DISTANCE;
let nodesPos = new Map(nodeID, nodePos);

function onNodePos(nodeID, nodePos)
	nodesPos.set(nodeID, pos);
	let nearbyNodeIDs = [];
	for each possiblePeerID in nodesPos.keys
		distance = calcDistance(possiblePeerID, nodeID)
		if(distance <= MAX_DISTANCE)
			nearbyNodeIDs.push(ID);
	client.send(nearbyNodeIDs);

\end{lstlisting}
\end{figure}


\begin{figure}
\begin{lstlisting}[language=bash, caption=Overlay client logic, label=lst:overlayclient]
const MAX_DISTANCE, MAX_PEERS, ANNOUNCEMENT_TIME
let lastSentPos = currentPos
let nodesOfInterest, peers = new Set()

function onServerConnection() 
	serverConnection.send(currentPos);
		
function onServerResponse(message) 
	nodesOfInterest.addAll(message.peerIDs);
	
function updatePosition(); //Every ANNOUNCEMENT_TIME ms
	if( calcDistance(lastSentPos, currentPos) > MAX_DISTANCE)
		if(!server)
			sendToPeers(currentPos, propagate=true, visitedPeers)
		else sendToPeers(currentPos, propagate=false); //TTL=1
		lastSentPos = currentPos;

function onPositionMessage(message)
	if(calcDistance(currentPos, message.currentPos) <= MAX_DISTANCE)
		nodesOfInterest.add(message.sender);
	if(message.propagate)
		message.visitedPeers.add(myNode); sendToPeers(message);
		
function reviewPeers()
	removeDistantPeers() // distance > MAX_DISTANCE
	// Until MAX_PEERS
	if(peerCount() < MAX_PEERS) sendPeerRequest(); 
	if(peerCount() > MAX_PEERS) removeMostDistantPeer();
\end{lstlisting}
\end{figure}

\textbf{\textit{The GeoLoc Objects Bully.}}
With the GeoLoc overlay algorithm, peers can establish connections amongst each other based on physical proximity. However, it is not the best practice to have all nodes directly connected with the object server, even in architectures with edge servers. To address this issue in our algorithm, a set of designated nodes in the overlay, named bullies, are responsible for communicating with the servers.
A naive approach would be to designate the node with the lowest ID in a region to be a bully - making all its peers disconnect from the object server.

While this protocol is more than sufficient when all clients are interested in all objects, it can result in undesirable situations when a client replicates only a tiny subset of all available objects. In such a situation, the node would have to forward objects' updates it is not interested in, or drop such updates. 


We devised an improved version of the naive bully algorithm, where nearby nodes interested in the same objects create specialised overlays among them (on top of the general GeoLoc overlay). The simplified logic is illustrated in Listing~\ref{lst:bully}. The idea is that, for every object a node is replicating, it finds a peer that is also interested in that object and holds an open link to an object server -- that peer is named as the bully for that specific object. Suppose a node is capable of doing this process to every object it holds an interest in. In that case, it means that a direct connection from that node to an object server is not needed, as it can rely on its peers to synchronise the information with a server.

This algorithm ensures that any node that executes operations on any object will send those changes to the object server. Suppose a node moves to a new location and encounters an object it wants to manipulate, although none of its peers holds a link to a server. Then the node itself will replicate the object from the server, performing the desired modifications and synchronising directly with the server, even if no other peer shows interest in that object.

\begin{figure}
\begin{lstlisting}[language=bash, caption=Object bully logic, label=lst:bully]
const BROADCAST_TIME, BULLY_TIMEOUT;
let bullies = new Map(objID, nodeID);
let objectsIDs = objectStore.objectsIDs;

function init()
	for each objID in objectsIDs
		bullies.set(objID, myNode.ID);
		
function broadcastBully() // Every BROADCAST_TIME ms
	for each objID in objectsIDs
		if(bullies.get(objID) == myNode.ID)
			broadcastPeersImTheBully(myNode.ID, objID);
			
function onImTheBullyMessage(peerID, objID)
	if(peerID <= bullies.get(objID))
		bullies.set(peerID, objID);
		setBullyTimeout(objID, BULLY_TIMEOUT);
	else
		if(bullies.get(objID) == myNode.ID && peerID > myNode.ID)
			sendImTheBully(myNode.ID, objID, peerID);

function onBullyTimeout(objID)
	bullies.set(objID, myNode.ID);
	
function onClientDisconnect(peer)
	for each objID in objectsIDs
		if(bullies.get(objID) == peer.ID)
			bullies.set(objID, myNode.ID);
\end{lstlisting}
\end{figure}


\section{Applications}

Several classes of applications might benefit from using our proposed algorithms on a large scale.
One group of applications are location-based massively multiplayer online games. Recent years have seen the release of dozens of those games (\cite{montola2009pervasive, magerkurth2005pervasive}). This proliferation is due to widespread mobile devices that handle multiple sensors, and an ever-rising compute power. It is only natural that developers think about employing those capacities to add new interaction capabilities to applications.

The other variety of applications we explore in this section is in the domain of location-based messaging. Most current social networks have chat features. Commonly, those apps rank at the peak of the top apps lists of the leading application stores, with a massive number of installations between the various mobile platforms.

\textbf{Pokémon Go Case Study.}
Pokémon GO (\cite{pokemongo}) is an augmented reality location-based game released in July 2016. Developed by Niantic, Inc. for Android and iOS devices, it uses the devices' GPS capabilities to discover, capture, train, and battle virtual creatures, called Pokémon.
Studying the game's mechanics and deployment, we ascertain that it relies heavily on Google Cloud infrastructure, including data storage and network stack. As stated in (\cite{beyer2018site}), all user actions and game state changes require a call to a data store. 
To improve system performance regarding aspects such as, reducing server load and the number of network packets travelling with client requests. We identified three key areas where our algorithms would help enhance the client's user experience when employed.

The location where Pokémons spawn is elected at random by an application server - it is believed that the type of terrain and position of nearby PokéStops biases those locations. That information needs to be shared with every client.
The client-server model states that the client will query the server within an interval to check if it needs to show the user close-at-hand Pokémons ready to be captured. Our algorithm changes the approach of this mechanic from having all clients contacting the server to only a few of them per a defined geographic region will send a request. Then those nodes will synchronise with their local peers so that every node knows about the spawning creature. This can be achieved by disseminating such information over the general GeoLoc overlay.

Let us think about the incorporation of edge nodes in the system. A more significant impact is expected as edge nodes may attend to a broader geographic region. Furthermore, being on the infrastructure side yields better backbone access to servers. Edge nodes free servers to accommodate more clients, and those clients profit from lower latency to access and manipulate data.

The second game mechanic we explore is when a player visits a PokéStop. A PokéStop is a virtual location in-game that is backed by a fixed place in the physical world. Those points can be as distinct as statues, popular venues where locals gather, public parks, libraries, to name a few. The density of PokéStops a player encounters will differ from region to region; they may visit as many as they like but need to wait at least five minutes before revisiting the same one.
We find this scenario interesting because we would like to see how our algorithms would affect interactions with only two parties - the client and the server. As these objects are of no interest to any clients' peers, our algorithm should act similarly to a client-server model, where we do not expect a drop in performance. However, information regarding these PokéStops can be owned by edge servers, as to reduce the load on the central infrastructure.

Lastly, it would be interesting to explore another case. The game allows players to battle their Pokémons against each other. A battle initiates in one of two flavours. First, a player can challenge other players that are in their surroundings. Alternatively, one may invite from their friend list a player located at any part of the globe.

The local battles are exciting for conducting research with small P2P clusters (a server and two clients interested in replicating data from such events). Since we provide a node discovery mechanism, applications should not need to implement their own -- all the required information is already flowing within the overlay. We think that by applying this tool, the server logic can be more straightforward, its computation reduced, and fewer messages need to cross the network. 
The second type of player's battles poses a different type of challenge. As players are further apart, the benefits from direct communications plus the use of edge resources fades. If each client owns a connection to the edge, we could see some speedup on data replication, leveraging the edge nodes' backbone network access.

\textbf{Telegram Case Study.}
In June 2019, Telegram announced a update (\cite{telegram_location_2019}), starring a couple of new features - Location-based Chats and Add People Nearby.
Those innovations are relevant, as they can be employed in places where large masses of people gather at a particular moment (sports and music events, political rallies, demonstrations) or day-to-day life(college campus, business and industrial centres, apartment complexes).

We pinpointed two particular aspects that could benefit from using the architecture and algorithms we introduce - complex connections to the infrastructure and improving user experience.
A mix of a lack of cell coverage and the absence of WiFi networks is a usual condition at events held at large stadiums or more disheartening as an aftermath of natural disasters. Both scenarios take an immense toll on infrastructure. 
In the first example, we expect a more recreational use of devices with a more substantial number of messages and larger files, as pictures and video are shared among event-goers. Users want content delivered instantly. Sending and receiving that data from a central location (a common bottleneck) presents enormous bandwidth and latency costs.
In the realms of crisis management and disaster recovery, the ability to communicate between local communities or local authorities is imperative. People may need to call for help, and organisations need to coordinate efforts efficiently. 

Our work enables an application to share data that only pertains to a given region. Since it is communication-medium agnostic, device-to-device connections are possible among nearby devices freeing needed resources to other mission-critical uses, both on infrastructure and cloud services. Such a scenario can be further improved with the deployment of edge assets, which guarantee that data is replicated between peers and at higher levels of hierarchy - edge nodes and, subsequently, the cloud.

\section{Evaluation}

We devised a series of experiments to evaluate the behaviour of the proposed algorithms in a scenario with node mobility. To create such a scenario, we implemented a naive version of the Foursquare City Guide application (\cite{foursquare_city}), with the functionalities of checking in and posting a review about a venue. 
We built our application on top of the Legion framework (\cite{legion}), which enables P2P replication for web and mobile nodes through the use of Delta-based CRDTs (\cite{van2016delta}).

\textbf{Application Model.}
Foursquare allows users to discover information on nearby locations, permitting them to perform a "Check-In" in a particular business and obtain a list of nearby places. Users may also use the "Explore" feature to search for a specific type of place using different categories and leave a review after a visit. 

To host our application, we introduced two modifications to the Legion codebase. 
First, we implemented the GeoLoc overlay and bully algorithms in Javascript, including the algorithms within the framework as a tool to propagate messages. Secondly, we introduced fine-grained partial replication since each node must replicate all information (within a large bucket) by default in Legion.

When a node sends data to one of its peers, one of two situations may happen. If the receiving end is an object server, then it is in its best interest to retain all received data. If the peer receiving data is a client, it needs to determine if its local store contains a version of that object.
If it does, the system synchronises both peers' states, ending with both nodes owning that object's same (reconciliated) version.

\textbf{Geographic dataset.}
To simulate the location of both clients and objects, we used two datasets. One contains the geographical coordinates of approximately 43 000 taxi trips within Porto in Portugal, which was enhanced with the help of the OpenRouteService API (\cite{neis2008openrouteservice}), resulting in a dataset with more coordinates per trip. The other dataset holds the coordinates of bus stops in that same city. We used the taxi trips dataset to simulate client movement, and the bus stops dataset for object locations.

\textbf{Experimental setup.}
We conceived scenarios with five clients moving through routes of forty positions each. The chosen routes have points close to each other to ensure that clients will be in the same area at some point.
For object locations, fifty positions were extracted. Considering that, as clients travel, they face objects on their paths, sets of clients share a subset of objects at some point in time, and a single client exclusively handles some objects.

We tested three different overlay models: employing the GeoLoc Overlay in partial replication mode (GLO-partial), in full replication mode (GLO-full), and for contrast, a simple client-server overlay in partial replication mode (C-S), which does not rely on any P2P communication.

\textbf{Experiment scenarios.}
The following descriptions are related to the use of a partial replication model. In the client-server overlay, a client in need of an object sends a request to the server directly, so only the server needs to hold all data. In the full replication model, each client holds all data. That reduces the need for a server but imposes more interaction between nodes since they need to exchange data to synchronise object versions.

\textbf{Check-in scenario.}
A check-in is performed at every venue of an area when a node moves to a new location. When approaching a venue, it needs to receive the object's current state representing that venue from nearby peers or a server. We use CRDT counters, as we only count the number of nodes that came close to a distinct spot, incrementing that counter by one unit. When a client moves away, the counter is removed from the client's local CRDT storage.

\begin{figure*}
	\centering
	\begin{minipage}{0.5\columnwidth}
		\centering
		\includegraphics[width=\textwidth]{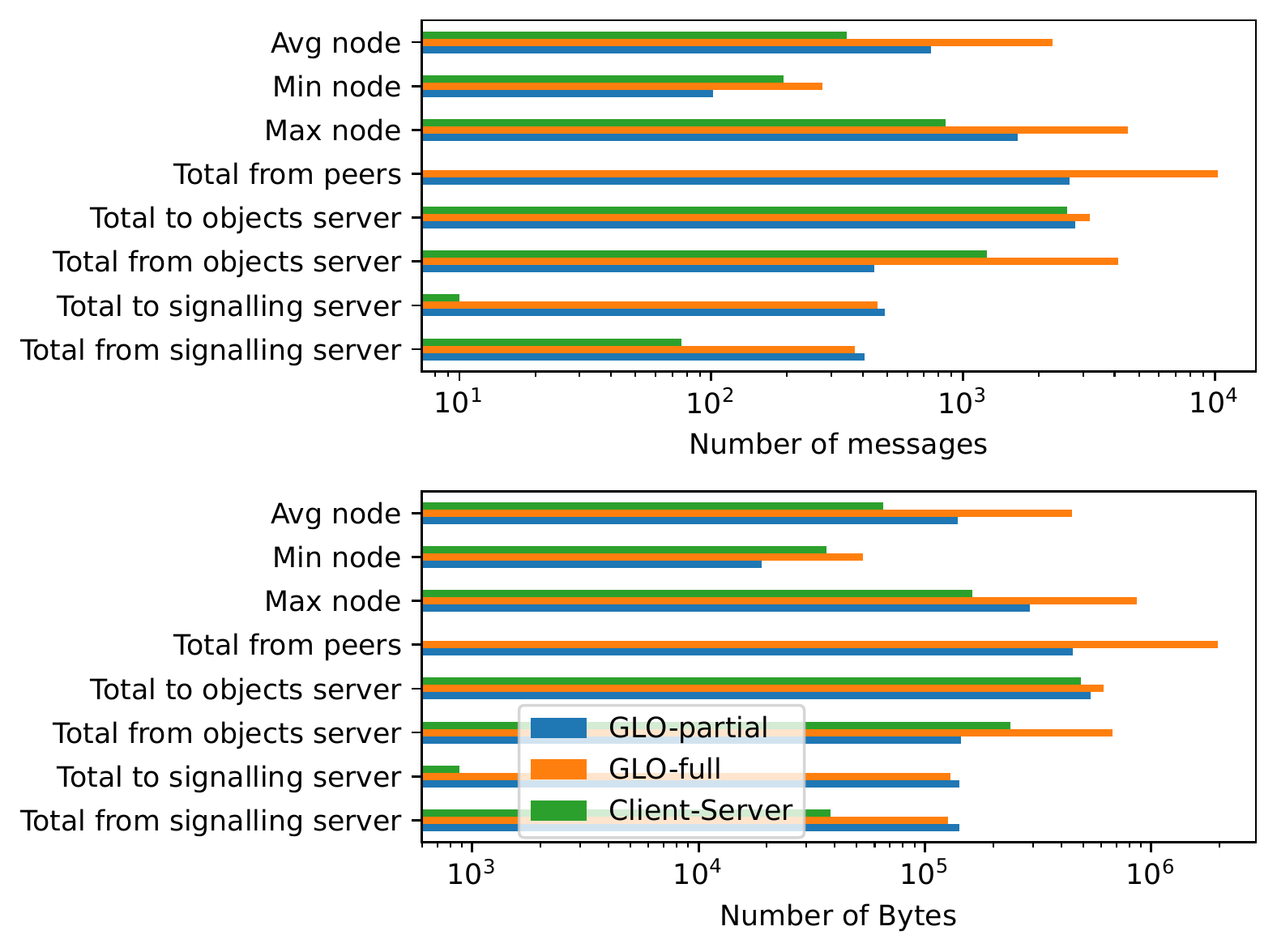}
		\caption{Check-in scenario}
		\label{fig:checkin}
	\end{minipage}%
	\begin{minipage}{0.5\columnwidth}
		\centering
		\includegraphics[width=\textwidth]{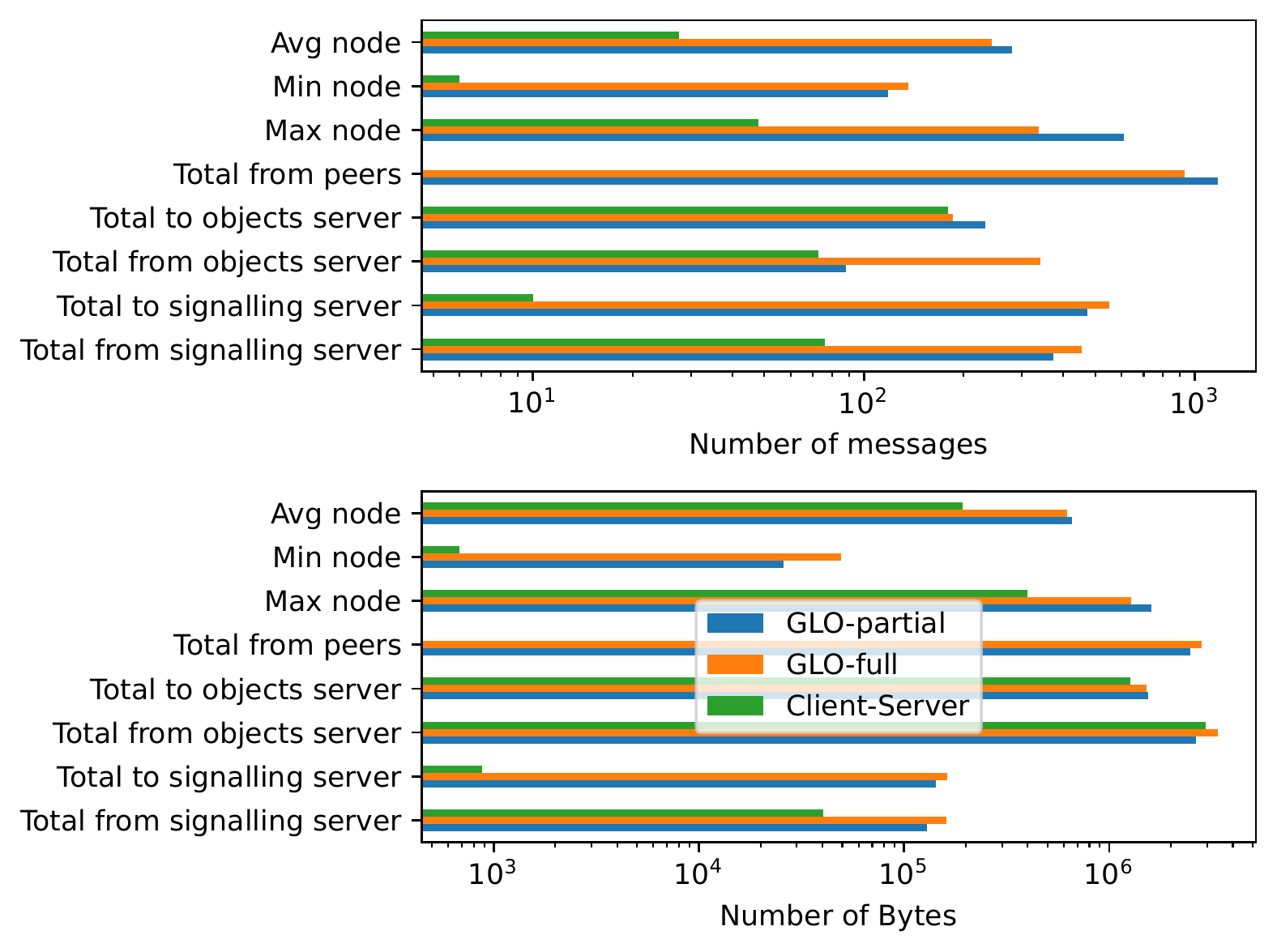}
		\caption{Post a Review scenario}
		\label{fig:review}
	\end{minipage}
\end{figure*}

The results from Figure~\ref{fig:checkin} show that using the GLO-partial generates an overall lower number of messages traded between nodes and objects servers and amongst peers when compared with GLO-full.
The C-S model still manages to send slightly fewer messages to the server when compared with GLO-partial. However, GLO-partial by allowing clients to receive only objects that they deem relevant, in fact, reduces the number of clients holding open connections to the server.

\textbf{Post a review scenario.}
We modelled the action of a client writing a review of a venue after a visit. When a client gets near a venue with some probability factor, it will generate a random string with a size varying from 500 bytes to 10kB. Then this string is stored on a CRDT map representing the venue. The process of synchronising and discarding object data from a node's local storage is performed similarly to the previous scenario. We could model other interactions with minor modifications, such as a threaded comment section.

From the results reported in Figure~\ref{fig:review}, we can observe that the number of messages exchanged between peers and sent from clients to the object server is higher in GLO-partial when compared with GLO-full. This occurs because, in this scenario, each node only interacts with an object once. A fair share of messages traded amongst peers is for overlay control, such as bullying and peer discovery. That fact consolidates by comparing the number of messages with their size using GLO-partial. Despite peers sending more messages among them, the total number of bytes sent is lower than that of the GLO-full model.

\textbf{Operations latency.}
To evaluate the latency of propagating operations on objects, we employed a single CRDT map shared among all clients. Every client applies a write to this map every ten seconds, letting this change propagate through the overlay. To measure latency, each write operation records its execution time, with the receiving node computing the delta to their receiving time. 

\begin{figure}[t]
\begin{center}
\includegraphics[width=.45\textwidth]{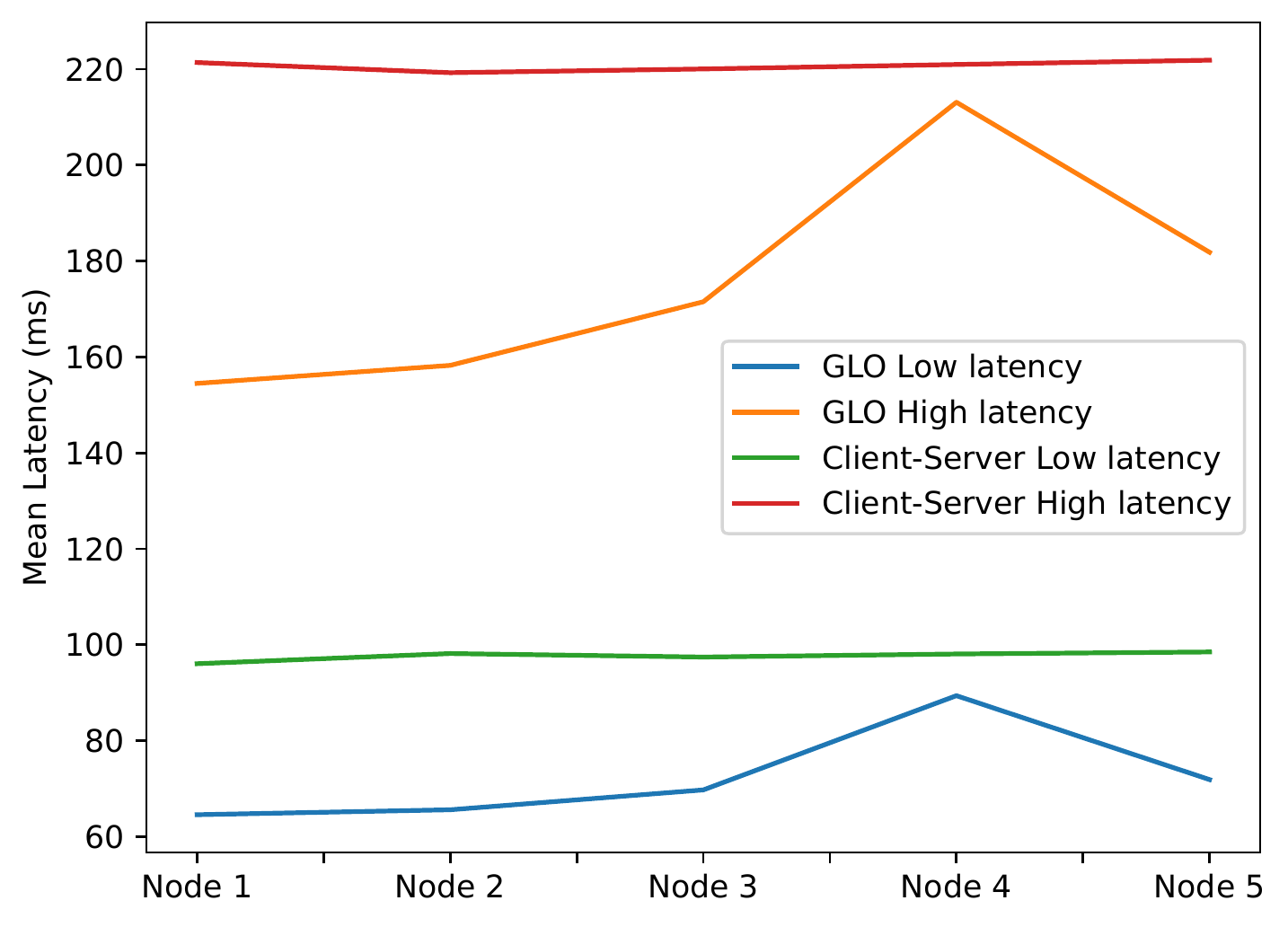}
\caption{Latency results}
\label{fig:latency}
\end{center}
\end{figure}

We performed this analysis in a private network, and we artificially introduced delays between peers. Making use of a mix of measured and reported latencies from several sources (\cite{farris2018providing, fiandrino2019openleon}).
The two latency classes - low and high - respectively derive from measurements among devices connected within the same infrastructure (i.e. P2P connections and to regional edge servers) and linking end devices and the cloud.
The results in Figure~\ref{fig:latency} show that using the GLO results in lower latency, explained by the fact that nearby nodes are capable of interacting directly without having to use the server as an intermediary.

\section{Conclusion and Future Work}
In this work, we presented a framework that enables clients to receive data that they hold an interest in, helped by monitoring each stakeholder's geographic position. Our newly developed overlay and bully algorithms provide clients with the ability to correctly and efficiently receive information on nearby objects. We are asserting a guarantee - if an object of interest does not propagate among nearby clients, a server assumes that responsibility by storing that data.

Our preliminary evaluation shows some overhead from the messages traversing the overlay with node position information. Unlike the full-replication model, our protocol proved to exchange more concise data (smaller in size and number). Compared with the client-server model, we achieved lower latency.

As for future work, we want to understand if we can decrease the values of numbers of messages in the overlay to near or lower than the client-server model with our protocol.
An extended evaluation is needed to achieve and further validate our concept. Such assessment ought to profit from a very large scale scenario with hundreds of thousands of clients. One possibility is to simulate the entire protocol stack.

\section*{Acknowledgments}
This work was partially supported by FCT/MCTES grants, through the SAMOA project  (PTDC/CCI-INF/32662/2017, Lisboa-01-0145-FEDER-032662), NOVA LINCS (UIDB/04516/2020) and the PhD Research Scholarships awarded to Albert van der Linde (SFRH/BD/117446/2016) and Luís Silva (2021.05686.BD).
	Dataset expansion was possible by API access to OpenRouteService from HeiGIT gGmbH.


\bibliographystyle{unsrtnat}
\bibliography{bibliography.bib}

\end{document}